\documentclass[12pt]{iopart}
\usepackage{amssymb}

\newcommand{\btr}{\bigtriangledown}

\newcommand{\la}{\lambda}
\newcommand{\lb}{\label}

\newcommand{\al}{\alpha}
\newcommand{\bt}{\beta}
\newcommand{\ka}{\kappa}

\newcommand{\ga}{\gamma}
\newcommand{\si}{\sigma}

\newcommand{\fr}{\frac}

\newcommand{\f}{{f}}
\newcommand{\bw}{\begin{widetext}}
\newcommand{\ew}{\end{widetext}}
\newcommand{\be}{\begin{equation}}
\newcommand{\ee}{\end{equation}}
\newcommand{\bee}{\begin{equation*}}
\newcommand{\eee}{\end{equation*}}
\newcommand{\ba}{\begin{eqnarray}}
\newcommand{\ea}{\end{eqnarray}}
\newcommand{\bal}{\begin{align}}
\newcommand{\eal}{\end{align}}
\newcommand{\bml}{\begin{multline}}
\newcommand{\eml}{\end{multline}}
\newcommand{\non}{\nonumber}

\def\e{{\rm e}}
\begin{document}
\title[Cerenkov radiation from collisions of straight cosmic
(super)strings] {CERENKOV RADIATION FROM COLLISIONS OF STRAIGHT
COSMIC (SUPER)STRINGS}

\author{ELENA MELKUMOVA, DMITRI GAL'TSOV and KARIM SALEHI}

\address{Department of Physics, Moscow State University,119899,
Moscow, Russia\footnote{Talk given at 11th Marcel Grossmann
Meeting On Recent Developments In Theoretical And Experimental
General Relativity, Gravitation, And Relativistic Field Theories,
23-29 Jul 2006, Berlin, Germany}} \ead{elenamelk@srd.sinp.msu.ru}

\begin{abstract}
We consider Cerenkov radiation which must arise when randomly
oriented straight  cosmic (super)strings move with relativistic
velocities without intercommutation. String interactions via
dilaton, two-form and gravity (gravity  being the dominant force
in the ultra-relativistic regime) leads to formation of
superluminal sources which generate Cerenkov radiation of dilatons
and axions. Though the effect is of the second order in the
couplings of strings to these fields, its total efficiency is
increased by high  dependence of the radiation rate on the
Lorentz-factor of the collision.
\end{abstract}
\section{Introduction}\label{intro}
Recently the early universe models involving strings and branes
moving in higher-dimensional space-times received a renewed
attention \cite{DuKu05}-\cite{DaKi05}. In particular, the problem
of the dimensionality of space-time can be explored within the
brane gas scenario \cite{DuKu05}-\cite{Po04}. Another new
suggestion is the possibility of cosmic superstrings with lower
tension than those in the field-theoretical GUT strings
\cite{Po04}. Superstrings as cosmic strings candidates revive the
idea of the defect origin of cosmic structures and stimulate
reconsideration of the cosmic string evolution with account for
new features  such as existence of the dilaton and antisymmetric
form fields and extra dimensions. The main  role in this evolution
is played by radiation processes. The radiation mechanism which
has been mostly studied in the past consists in formation of the
excited closed loops  which subsequently loose their excitation
energy emitting gravitons \cite{VaVi85} axions \cite{DaQu89} and
dilatons \cite{Da85}-\cite{ViVa85}.

In this paper we consider the bremsstrahlung mechanism of string
radiation  \cite{GaGrLe93} which works for initially unexcited
strings undergoing a collision. We develop a classical
perturbation scheme for two endless unexcited long strings which
move one with respect to another in two parallel planes being
inclined at an angle. It was shown earlier that in four space-time
dimensions there is no gravitational bremsstrahlung under
collision of straight strings \cite{GaGrLe93}. This can be traced
to absence of gravitons in 1+2 gravity. It is not a coincidence
that in four dimensions there is no gravitational renormalization
of the string tension either \cite{BuDa98}. But there is no such
dimensional argument in the case of the axion field there  such
dimensional argument and it was demonstrated that string
bremsstrahlung takes place indeed \cite{GMK04} within the model in
flat space. Here we extend this result to the full gravitating
case including also the dilaton field. Strings interacts via the
dilaton, axion and graviton exchange. Radiation arises in the
second order approximation in the coupling constants provided the
(projected) intersection point moves with superluminal velocity.
Thus, the string bremsstrahlung can be viewed as manifestation of
the Cherenkov effect.

\section{ String interactions}
Consider a pair of relativistic strings
$x^{\mu}=x_n^{\mu}(\sigma^a_n),\;\; \mu=0,1,2,3,\;\;
\sigma_a=(\tau, \sigma),\;\; a=0,1,$ where  $n=1,2$ is the  index
labelling the two strings. The 4-dimensional space-time metric
signature $+,---$ and $(+,-)$ for the string world-sheets metric
signature. Strings interact via the gravitational $ g_{\mu\nu}
\equiv \eta_{\mu\nu} + h_{\mu\nu} $, dilatonic $ \phi (x) $ and
axion (Kalb-Ramond) field $B_{\mu\nu}(x)$: \begin{eqnarray} S
=-\sum_n\int\left\{
 \frac{\mu}{2} \partial_a x^\mu_n \partial_b x^\nu_n
 g _{\mu\nu}\gamma^{ab}\sqrt{-\gamma}\exp^{2\alpha_n\phi}+2\pi\f
  \partial_a x^\mu_n \partial_b x^\nu_n \epsilon^{ab}
  B_{\mu\nu}\right\}d^2 \sigma \non\\+
\int\limits\left\{ 2\partial_\mu\phi\partial_\nu\phi
g^{\mu\nu}+\fr{1}{6}H_{ \mu \nu \rho} H^{ \mu \nu
\rho}\e^{-4\al\phi} - \frac{R}{16\pi G } \right\}\sqrt{-g} \,
d^4x.\label{fgac} \end{eqnarray}
 Here $\mu_n$ are the (bare) string tension
parameters, $\al$ and $f$ are the corresponding coupling
parameters, $\epsilon^{01}=1,\; \ga_{ab}$ is the induced metric on
the world-sheets. In what follows, we linearize the dilaton
exponent as $ \e^{2\al\phi}\simeq 1+2\al\phi $.
\par
The totally antisymmetric axion field strength is defined as $H_{
\mu \nu \la}=\partial_\mu B_{\nu\la}+\partial_\nu
B_{\la\mu}+\partial_\la B_{\mu\nu}.$ Variation of the action
(\ref{fgac}) over $ x^\mu_n $ leads to the  equations of motion
for strings \ba \partial_a\left(\mu
\partial_b x^\nu_n
g_{\mu\nu}\ga^{ab}\sqrt{-\ga}\e^{2\al\phi}+4\pi f
\partial_b x^\nu_n\epsilon^{ab}B_{\mu\nu}\right)\non\\-
\mu\al \partial_a x^\al_n
\partial_b x^\beta_n
g_{\al\beta}\ga^{ab}\sqrt{-\ga}
\e^{2\al\phi}\partial_{\mu}\phi-\fr\mu2 \partial_a x^\al_n
\partial_b x^\beta_n\ga^{ab}\sqrt{-\ga}
\e^{2\al\phi}\partial_\mu g_{\al\beta}=0.\label{steq}\ea Variation
with respect to field variables $ \phi $, $ B_{\mu\nu} $ and $
g_{\mu\nu} $ leads to the dilaton equation: \be
\partial_\mu\left(g^{\mu\nu}\partial_\nu
\phi\sqrt{-g}\right)+\fr{\al}6
H^2\e^{-4\al\phi}+\fr{\mu\al}{4}\int
\partial_a x^\mu_n
\partial_b x^\mu_n
g_{\mu\nu}\ga^{ab}\e^{2\al\phi}\delta^4(x-x_n(\si_n))d^2\sigma=0,\label{fep}\ee
the axion  equation: \be\label{feB}
\partial_\mu\left(H^{\mu\nu\la}\e^{-4\al\phi}\sqrt{-g}\right)+2\pi
f \int \partial_a x^\nu_n \partial_b x^\la_n
\epsilon^{ab}\delta^4(x-x_n(\si_n))d^2\sigma=0\ee and the Einstein
equations:  $ R_{\mu\nu}-\fr12g_{\mu\nu}R=8\pi
G(\stackrel{\phi}{T}{\!}_{\mu\nu}+\stackrel{B}{T}{\!}_{\mu\nu}+
\stackrel{st}{T}{\!}_{\mu\nu}),$ \\
$\stackrel{st}{T}{\!}_{\mu\nu}=\sum \mu\int
\partial_a x_{\mu \, n}
\partial_b x_{\nu \,
n}\gamma^{ab}\sqrt{-\gamma}\e^{2\al\phi}
\fr{\delta^4(x-x_n(\si_n))}{\sqrt{-g}}d^2\sigma,$

$\stackrel{\phi}{T}{\!}_{\mu\nu}\!=\!4\left(\partial_\mu\phi\partial_\nu\phi-\fr12
g_{\mu\nu}(\btr\phi)^2\right),$ $
\stackrel{B}{T}{\!}_{\mu\nu}=\left(H_{\mu\al\bt}H_\nu^{\al\bt}-
\fr16H^2g_{\mu\nu}\right)\e^{-4\al\phi}.$
\par
Our calculation follows the approach of
\cite{GaGrLe93}-\cite{GMK04} and  consists in constructing
solutions of the string equations of motion and dilaton, axion and
graviton iteratively using the coupling constants $\alpha,\,f,\,G$
as expansion parameters.
\par
The total  dilaton, axion and graviton fields are the sums due to
contributions of two strings: $\phi=\phi_1+\phi_2,\;
B^{\mu\nu}=B^{\mu\nu}_1+B^{\mu\nu}_2,\;
h^{\mu\nu}=h^{\mu\nu}_1+h^{\mu\nu}_2.$ Since in the zero order the
strings are moving freely , the first order dilaton $
\stackrel{1}{\phi} _{n} $, axion $\stackrel{1}{B}{\!}^{\mu\nu}_n $
and graviton variables $\stackrel{1}{h}{\!}^{\mu\nu}_n $  do not
contain radiative components. Substituting them into the Eq.
(\ref{steq}) we then obtain the first order deformations of the
world-sheets $\stackrel{1}{x}{\!}^{\mu} $, which are naturally
split into contributions due to dilaton, axion and graviton
exchange: $ \stackrel{1}{x}{\!}^{\mu}_n =
\stackrel{1}{x}{\!}^{\mu}_{n (\phi)} +
\stackrel{1}{x}{\!}^{\mu}_{n (B)}+\stackrel{1}{x}{\!}^{\mu}_{n
(h)}.$
\par
Radiation arises in the second order field terms
${\stackrel{2}\phi}_{n}$ and ${\stackrel{2}B}{\!}^{\mu\nu}_n$
which are generated by the first order currents
$\stackrel{1}J_{(\phi)}$,${\stackrel{1}J}{\!}^{\,\mu\nu}_{(B)}$
 in the dilaton and axion field equations
((\ref{fep}),(\ref{feB})).

  Note that gravitational radiation in four dimensions  is
absent \cite{GaGrLe93}, so we do not consider the second order
graviton equation.  The dilaton and axion radiation power can be
computed as the reaction work given by the half sum of the
retarded and advanced fields upon the sources \cite{GMK04}.  The
final formula for the dilaton and axion bremsstrahlung from the
collision of two global strings can be obtained analytically in
the case of the
ultrarelativistic collision with the Lorentz factor $\gamma=(1-v^{2}%
)^{-1/2}\gg1$. We assume the BPS condition for  the coupling
constants \cite{BuDa98} $\al\mu=2\sqrt{2}\pi f$. The main
contribution to radiation turns out to come from the graviton
exchange terms. The spectrum has an infrared divergence due to the
logarithmic dependence of the string interaction potential on
distance, so a cutoff length $\Delta$ has to be introduced: \be
\fl P^{(\phi)}=\fr{200}{3}\pi G^2\al^2\mu^4
L\ka^5(f(y)+\fr1{25}f_1(y)),\quad P^{(B)}=\fr{16\pi^3 G^2\mu^2L
f^2\kappa^5}{3 }(f(y)-f_2(y)),\lb{Totp1}\ee where $L$-length of
the string, $y=\frac{d}{\gamma\kappa\Delta}$,
$\kappa=\gamma\cos\alpha $, $\alpha$ is the strings inclination
angle, $d$ is the impact parameter and
$f(y)=12\sqrt{\fr{y}{\pi}}\,
{_2F_2}\left(\fr12,\fr12;\fr32,\fr32;-y\right)
-3\ln\left(4y\e^{C}\right),$ $ f_1(y)={\rm
erfc}(\sqrt{y})\left(\fr83y^3-30y^2+114y+\fr{169}{2}\right)-\fr{\e^{-y}\sqrt{y}}{\pi}\left(
\fr{8}3y^2-\fr{94}{3}y+131 \right), $ $ f_2(y)={\rm
erfc}(\sqrt{y})\left(\fr83y^3+6y^2-6y-\fr{5}{2}\right)-\fr{\e^{-y}\sqrt{y}}{\pi}\left(
\fr83y^2+\fr{14}{3}y-7 \right), $ $F$ is the generalized
hypergeometric function  and $C$ is the Euler's constant.\par
This work  was  supported in part by RFBR grant 02-04-16949.

\vfill

\end{document}